# Roadmap for Molecular Benchmarks in Nonadiabatic Dynamics

**Léon L. E. Cigrang**†,[1] **Basile F. E. Curchod**†,[2] **Rebecca A. Ingle**†,[1] **Aaron Kelly**†,[3] **Jonathan R. Mannouch**†,[3] **Davide Accomasso**,[4,5] **Alexander Alijah**,[6] **Mario Barbatti**,[7,8] **Wiem Chebbi**,[9] **Nađa Došlić**,[10] **Elliot C. Eklund**,[11] **Sebastian Fernandez-Alberti**,[12] **Antonia Freibert**,[13] **Leticia González**,[14] **Giovanni Granucci**,[15] **Federico J. Hernández**,[2] **Javier Hernández-Rodríguez**,[16] **Amber Jain**,[17] **Jiří Janoš**,[18,2] **Ivan Kassal**,[11] **Adam Kirrander**,[19] **Zhenggang Lan**,[20,21] **Henrik R. Larsson**,[22] **David Lauvergnat**,[23] **Brieuc Le Dé**,[24] **Yeha Lee**,[25] **Neepa T. Maitra**,[26] **Seung Kyu Min**,[27] **Daniel Peláez**,[28] **David Picconi**,[29] **Zixing Qiu**,[28,30] **Umberto Raucci**,[31] **Patrick Robertson**,[32] **Eduarda Sangiogo Gil**,[33] **Marin Sapunar**,[10] **Peter Schürger**,[23] **Patrick Sinnott**,[11] **Sergei Tretiak**,[34] **Arkin Tikku**,[11] **Patricia Vindel-Zandbergen**,[35,36] **Graham A. Worth**,[1] **Federica Agostini***,[23] **Sandra Gómez***,[37] **Lea M. Ibele***,[7,23] and **Antonio Prlj***[10]

[1]*Department of Chemistry, University College London,*
*20 Gordon St., WC1H 0AJ London, United Kingdom*

[2]*Centre for Computational Chemistry, School of Chemistry,*
*University of Bristol, Bristol BS8 1TS, United Kingdom*

[3]*Hamburg Center for Ultrafast Imaging,*
*Universität Hamburg and Max Planck Institute for the*
*Structure and Dynamics of Matter, 22761 Hamburg, Germany*

[4]*Faculty of Chemistry, University of Warsaw, Pasteura 1, Warsaw, Poland*

[5]*Department of Industrial Chemistry, University of Bologna, via Gobetti 85, Bologna, Italy*

[6]*Groupe de Spectrométrie Moléculaire et Atmosphérique, GSMA,*
*UMR CNRS 7331, Université de Reims Champagne-Ardenne,*
*U.F.R. Sciences Exactes et Naturelles,*
*Moulin de la Housse B.P. 1039, 51687 Reims Cedex 2, France*

[7]*Aix Marseille University, CNRS, ICR, 13397 Marseille, France*

[8]*Institut Universitaire de France, 75231 Paris, France*

[9]*Laboratoire de Spectroscopie Atomique, Moléculaire et Applications (LSAMA),*
*University of Tunis El Manar, 1060 Tunis, Tunisia*

[10]*Department of Physical Chemistry, Ruđer Bošković Institute,*

*Bijenička cesta 54, 10000 Zagreb, Croatia*

[11]*School of Chemistry, University of Sydney, NSW 2006, Australia*

[12]*Departamento de Ciencia y Tecnologia, Universidad Nacional de Quilmes/CONICET, B1876BXD Bernal, Argentina*

[13]*Department of Physics, University of Hamburg, Luruper Chaussee 149, 22761 Hamburg, Germany*

[14]*Institute of Theoretical Chemistry, Faculty of Chemistry, University of Vienna, Währinger Str. 17, A-1090 Vienna, Austria*

[15]*Department of Chemistry and Industrial Chemistry, University of Pisa, via Moruzzi 13, 56124 Pisa, Italy*

[16]*Departamento de Química Física, Universidad de Salamanca, Salamanca 37008, Spain*

[17]*Department of Chemistry, Indian Institute of Technology Bombay, Mumbai 400076, India*

[18]*Department of Physical Chemistry, University of Chemistry and Technology, Technická 5, 16628 Prague, Czech Republic*

[19]*Physical and Theoretical Chemistry Laboratory, Department of Chemistry, University of Oxford, South Parks Road, OX1 3QZ Oxford, United Kingdom*

[20]*MOE Key Laboratory of Environmental Theoretical Chemistry, South China Normal University, Guangzhou 510006, China*

[21]*SCNU Environmental Research Institute, Guangdong Provincial Key Laboratory of Chemical Pollution and Environmental*

[22]*Department of Chemistry and Biochemistry, University of California, Merced, CA 95343, USA*

[23]*Université Paris-Saclay, CNRS, Institut de Chimie Physique UMR 8000, 91405 Orsay, France*

[24]*Sorbonne Université, CNRS, Institut des Nanosciences de Paris, 75005 Paris, France*

[25]*Laboratory of Theoretical Physical Chemistry, Institut des Sciences et Ingénierie Chimiques, Ecole Polytechnique Fédérale de Lausanne (EPFL), CH-1015 Lausanne, Switzerland*

[26]*Department of Physics, Rutgers University, Newark, New Jersey 07102, USA*

[27]*Department of Chemistry, Ulsan National Institute of Science and Technology (UNIST), South Korea*

[28]*Université Paris-Saclay, CNRS, Institut des Sciences Moléculaires d'Orsay, 91405, Orsay, France*

[29] *Institute of Theoretical and Computational Chemistry, Heinrich-Heine-Universität Düsseldorf, Universitätstraße 1, 40225 Düsseldorf, Germany*
[30] *MICS, CentraleSupélec, Paris-Saclay University, Gif-sur-Yvette, France*
[31] *Italian Institute of Technology, Via Enrico Melen 83, Genoa 16153, Italy*
[32] *School of Chemistry, University of Nottingham, Nottingham, NG72RD, United Kingdom*
[33] *Institute of Theoretical Chemistry, Faculty of Chemistry, University of Vienna, Währinger Str. 17, 1090 Vienna, Austria*
[34] *Theoretical Division and Center for Integrated Nanotechnologies, Los Alamos National Laboratory, Los Alamos, NM 87545, USA*
[35] *Department of Chemistry, New York University, New York, New York 10003, USA*
[36] *Simons Center for Computational Physical Chemistry at New York University*
[37] *Departamento de Química, Módulo 13, Universidad Autónoma de Madrid, Cantoblanco 28049 Madrid, Spain* [*]

ABSTRACT: Simulating the coupled electronic and nuclear response of a molecule to light excitation requires the application of nonadiabatic molecular dynamics. However, when faced with a specific photophysical or photochemical problem, selecting the most suitable theoretical approach from the wide array of available techniques is not a trivial task. The challenge is further complicated by the lack of systematic method comparisons and rigorous testing on realistic molecular systems. This absence of comprehensive molecular benchmarks remains a major obstacle to advances within the field of nonadiabatic molecular dynamics. A CECAM workshop, *Standardizing Nonadiabatic Dynamics: Towards Common Benchmarks*, was held in May 2024 to address this issue. This Perspective highlights the key challenges identified during the workshop in defining molecular benchmarks for nonadiabatic dynamics. Specifically, this work outlines some preliminary observations on essential components needed for simulations and proposes a roadmap aiming to establish, as an ultimate goal, a community-driven, standardized molecular benchmark set.

† L. L. E. C., B. F. E. C., R. A. I., A. K. and J. R. M. contributed equally to this work.

* sandra.gomezr@uam.es; federica.agostini@universite-paris-saclay.fr; lea-maria.ibele@univ-amu.fr; antonio.prlj@irb.hr

## I. INTRODUCTION

Modeling the dynamical behavior of a molecular system upon photoexcitation is a formidable theoretical and computational challenge. This is due to the involved coupled electron-nuclear dynamics, the so-called nonadiabatic effects, that necessitate treatment beyond the Born-Oppenheimer approximation.[1–4] As a result, the development of methods for simulating nonadiabatic molecular dynamics (NAMD) remains a key area of focus,[5–9] with research groups in theoretical chemistry and chemical physics having been particularly active in improving and testing simulation methods for several decades.

The field of NAMD has benefited from advances in experimental techniques capable of imaging the coupled electron-nuclear dynamics of molecules upon light absorption. The development of ultrashort laser pulses and subsequent experiments in femtochemistry[10] revealed direct measurements of dynamical processes in molecules,[11–15] which have helped validating and guiding advances in NAMD methodologies. However, the field of NAMD has also been fueled by significant progress in the development of efficient algorithms and software[16–31] able to solve the coupled dynamics of electrons and nuclei. Thanks to all of these developments, strong collaborations between experiments and theory, for example in the context of gas-phase time-resolved spectroscopy, have emerged for studying the excited-state dynamics of many systems, and have sparked community efforts aiming to challenge the predictive power of various NAMD methods.[32] Despite these significant steps toward unraveling the ultrafast photodynamics of numerous molecular systems, the field of NAMD still faces considerable challenges, as the following sections will discuss. At the same time, the reliability and trustworthiness of theoretical predictions often remain difficult to assess.

As a result, the community of developers and users of NAMD has recently pointed out the critical need for established and standardized benchmarks to advance the field. By *benchmark* here, we mean well-chosen systems that can be used to compare and test computational methods, along with a community-accepted robust procedure to be followed whenever newly developed methods and approximations in NAMD are tested. Benchmarks in NAMD are needed to improve methodologies, ensure reproducibility, estimate the reliability of the predictions, and enable theoretical developments to keep pace with experimental techniques.

Further, we believe that benchmarking existing methodologies and codes will ultimately assist users and newcomers to NAMD in identifying the most suitable technique for addressing specific problems.

Benchmarking has long been a cornerstone of computational chemistry, for instance of electronic structure theory,[33–36] and computational sciences in general.[37, 38] While such efforts serve as inspiring examples, NAMD presents unique challenges due to its inherent complexity. The outcome of an NAMD simulation relies on the calculation of observables with intricate time and energy dependencies, which are often associated with a wide range of physical and chemical phenomena. The high dimensionality of realistic molecular systems makes it impossible to simulate their quantum dynamics exactly. As a result, using approximations to the time-dependent molecular Schrödinger equation is necessary, which has driven the development of numerous NAMD methods over the past 40 years.

A CECAM workshop,[39] entitled *Standardizing nonadiabatic dynamics: towards common benchmarks*, took place in Paris in May 2024 with the central goal of stimulating the NAMD community towards developing a common benchmark set by (i) agreeing on the main ingredients required to test all families of NAMD techniques, of which we will provide examples in Section II, and (ii) selecting potential molecular systems for further tests. This Perspective summarizes the main conclusions reached during the CECAM workshop, aiming to inform the broader scientific community and encourage future benchmark efforts. More specifically, this Perspective serves as an opportunity to elaborate on key questions that emerged from the workshop regarding what makes a proper benchmark in NAMD.

Discussions made it clear that, given the complexity of NAMD simulations, initial attempts to propose realistic molecular benchmarks should begin with simple systems, namely small or medium-sized molecules in the gas phase. Even with such a limited focus, numerous open questions still arose during the discussions in the workshop.

- What constitutes an adequate reference for a benchmark in NAMD? An experiment or an accurate simulation?

- How do we decide which observables should be prioritized when establishing the reliability of a given method?

- How can NAMD methods based on fundamentally different theoretical frameworks be compared, such as those based on wavefunctions and those based on trajectories?

- How can different electronic-state representations and the intricacies of electronic-structure methods be handled?

- How can we even ensure that different NAMD techniques are initialized in the same way for a given benchmark system?

- How can we ensure that statistical convergence of computational results is achieved?

In addition to offering a structure for this Perspective article, the questions above highlight key topics that require dedicated attention to ensure the definition of proper and generalized benchmark systems in nonadiabatic dynamics. Accordingly, Section II A proposes some prototypical phenomena and related families of molecular systems that were considered appropriate for benchmarking. Section II B is dedicated to the different families of NAMD methods, aiming to identify the most representative theoretical approaches that can be used for a systematic comparison. Section II C provides a brief overview of the issues related to various electronic-structure methods for obtaining electronic energies and other electronic properties. In Section II D, we discuss the problem of the initial conditions for NAMD and how to ensure an equivalent initialization of the dynamics across different theoretical methodologies. Section II E identifies suitable physical observables and properties that can be directly calculated in an NAMD simulation and used in the context of benchmarking. In Section II F, we examine the role of experimental measurements and their suitability as a reference for NAMD. Finally, Section III summarizes the key insights that emerged from the CECAM workshop and outlines practical strategies for the community to advance the initiative of establishing robust benchmark systems for NAMD. We discuss how members of the community with diverse expertise can organize, share data, and collaborate effectively while also briefly exploring the future prospects for benchmarking. In this sense, this Perspective acts as a *roadmap* for future developments in NAMD.

## II. TOWARDS MOLECULAR BENCHMARKS: GENERAL CONSIDERATIONS

So far, mostly low-dimensional models of nonadiabatic processes have served as benchmark sets for NAMD[40–46], while many software packages try to incorporate such models in their benchmarking capabilities.[19, 47, 48] These models have often been engineered to challenge specific aspects of the NAMD formalism and offer the great advantage that they typically have numerically exact results to compare with. The famous Tully models, proposed in 1990 to evaluate the accuracy of the trajectory surface hopping method,[40] are still nowadays widely used by the NAMD community. This set of three one-dimensional model systems was specifically designed to investigate prototypical nonadiabatic processes, including single and multiple nonadiabatic crossings. However, the overarching goal of NAMD methods is to describe the photodynamics of a molecule in its full dimensionality. Thus, benchmarking NAMD on realistic photochemical processes is necessary to provide justification for their suitability in the simulations of the molecular systems of interest. In addition, while some approximations are thoroughly tested and understood for low-dimensional problems, their performance in higher dimensions is not necessarily known. In this respect, multidimensional model potentials are also important for benchmarking, and are often used by the community especially to compare fully-quantized and mixed quantum-classical approaches.[42, 49–51]

Constructing multidimensional potential energy surfaces (PESs) can rapidly become a very difficult task. The situation becomes even more complex for large molecular systems (hundreds of atoms) involving dozens of coupled excited states, where energies, gradients, and nonadiabatic couplings must all be considered. In these cases, strategies to reduce computational costs, such as dynamically limiting the number of excited states or nonadiabatic couplings to be calculated, become indispensable.[52]

The challenge of constructing multidimensional PESs is often circumvented by performing excited-state dynamics based on on-the-fly electronic structure calculations. Therefore, the concept of benchmarks needs to be adapted to on-the-fly NAMD, as was recently done with the "molecular Tully models", composed of ethylene, 4-$N$,$N'$-dimethylaminobenzonitrile (DMABN), and fulvene.[53] This benchmark set has been adopted by the community, and already several NAMD methods have been tested on one or more of these systems.[54–61] While useful, the molecular Tully models have shortcomings, such as the limited set of properties that have been used for comparisons, the fact that initial conditions were oriented towards trajectory-based methods, and the fact that only commercial software has been used for the underlying electronic structure, preventing broader accessibility and reproducibility. Even leaving the electronic-structure problem aside, it is evident that developing generalized and reliable benchmark sets for NAMD comes as a stringent challenge and currently hampers further developments in the field.

### A. Selected photophysical and photochemical phenomena for benchmarking

This Section describes photophysical and photochemical processes that could be used to assess the performance of different NAMD methods. In the following, we propose to select a few specific light-triggered phenomena, with the aim of achieving two main goals: narrowing down the choice of current benchmark systems, and providing some clear points of comparison between the results of different NAMD calculations (see also Section II E). The chosen phenomena should cover diverse aspects of photodynamics, highlighting the role of both nuclear and electronic effects. They should be generally well understood to avoid controversies related to the interpretation of the results. Additionally, we choose to privilege unimolecular processes in order to avoid unnecessary complexities in the early stage of building a benchmark strategy. With these elements in mind, the following four types of processes, which are briefly described below and illustrated in Figure 1, were preselected as interesting test systems.

- Photoisomerization (ISO): $\mathrm{ABC} + \mathrm{h}\nu \rightarrow \mathrm{CAB}$
- Photodissociation (DIS): $\mathrm{AB} + \mathrm{h}\nu \rightarrow \mathrm{A} + \mathrm{B}$
- Nonreactive radiationless relaxation (NRR): ${}^1\mathrm{A} + \mathrm{h}\nu \rightarrow {}^1\mathrm{A}^*[\rightarrow {}^3\mathrm{A}^*] \rightarrow {}^1\mathrm{A}$
- Excited-state intramolecular proton transfer (ESIPT): $\mathrm{AH} \cdots \mathrm{B} + \mathrm{h}\nu \rightarrow \mathrm{A} \cdots \mathrm{HB}$.

We note, however, that this list of phenomena is by no means exhaustive and future benchmark efforts will extend this selection to include, for instance, systems with high densities of electronic states, initialized in a coherent superposition of states, systems undergoing photoinduced electron transfer, or excited-state energy transfer or charge migration, or complex dynamical processes such as molecular collisions.

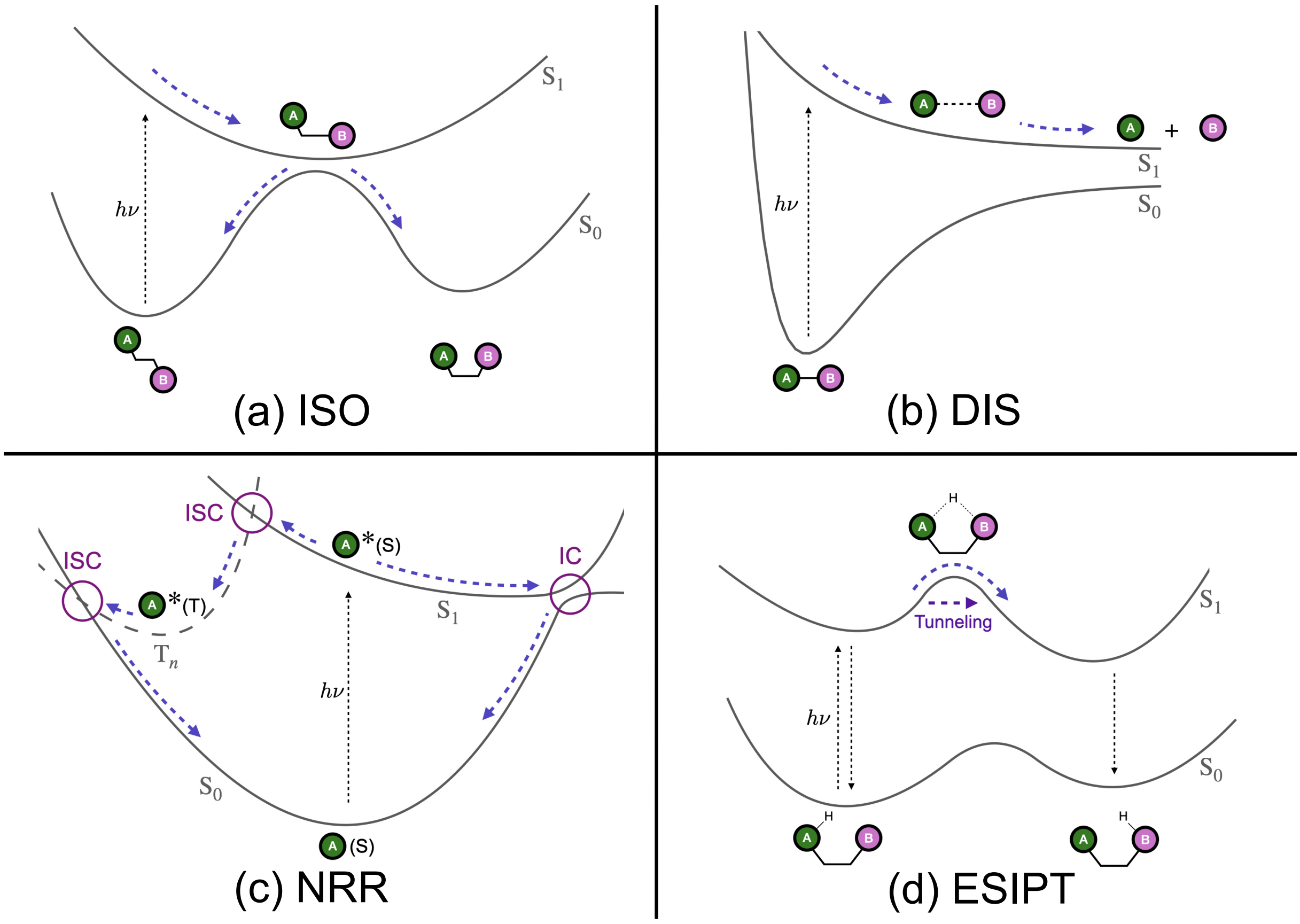

FIG. 1: Schematic representation of the four phenomena of interest for the benchmarking of NAMD methods: (a) Photoisomerization, i.e. an isomerization induced by photoexcitation; (b) Photodissociation, i.e. bond-breaking activated by absorption of light; (c) Nonreactive radiationless relaxation, i.e. transition between electronic states initiated by photoabsorption, potentially involving different spin multiplicities, without resulting in different photoproducts; and (d) Excited-state intramolecular proton transfer, occurring upon photoexcitation, typically involving hydrogen-bonded donor and acceptor groups. The curves labeled with S and T represent singlet and triplet potential energy curves, respectively, and the asterisk indicates an excited state of a species. The blue dashed arrows show nuclear motion along a molecular coordinate.

The above phenomena are some of the most common photoinduced processes in molecules

(see textbooks on molecular photophysics and/or photochemistry[2, 62–65]) and are representative of many of the current applications of NAMD methods. A search through the Semantic Scholar Academic Graph[66] indeed reveals that, when using the keywords "nonadiabatic dynamics", at least 50% of research articles on molecular systems published in the last 10 years discuss one of the four phenomena outlined above. Each of these phenomena presents a different challenge from a theoretical perspective: complex interplay between electronic character and nuclear motion (DIS), potential involvement of tunneling effects (ESIPT), molecular rearrangement (ISO), or complex transfer of electronic population (NRR). In the following, we discuss a few selected examples (i.e., not an exhaustive list) that demonstrate the importance of such light-induced processes.

One of the most paradigmatic examples of ISO is the cis → trans isomerization of retinal induced by photon absorption in mammalian eyes.[67, 68] Azobenzenes, stilbenes, and spiropyrans are also prominent classes of compounds subject to photoisomerization. [69–71] As shown in Figure 1a, photoisomerization[72] begins with light absorption, which weakens an originally locked bond (for example, by promoting an electron from a $\pi$ to a $\pi^*$ molecular orbital in a $\pi$ bond). This weakening of the originally locked bond allows the molecule to rearrange easily, often by rotation around a (pseudo-)single bond. In many photoisomerization processes, such as those involving retinal, azobenzene, and stilbene, the first excited state possesses a minimum located very close to the $S_0/S_1$ conical intersection (see Figure 1a). This minimum often corresponds to a geometry where the isomerization dihedral angle is close to 90–110°. Upon reaching this region of configuration space, the molecule can either proceed with a full photoisomerization or return to the original isomer, typically through a nonradiative decay involving a conical intersection and additional nuclear motion.[72]

In the field of femtochemistry, DIS was one of the first studied processes,[10, 73] and is schematically represented in Figure 1b. In the series of pioneering experiments by Zewail, the study of the photodissociation of iodocyanide (ICN)[74] preceded that of sodium iodide (NaI),[75] which was already studied in the earliest experiments of Polanyi.[76] Wavepacket dynamics simulations have supported these experiments from the beginning[43, 77–81] and NaI, as well as similar alkalihalides, has been extensively used as a simple one-dimensional test case for quantum dynamics methods since.[82–84] Absorption of UV light by small organic molecules such as, for example, methanol, phenol or pyrrole readily leads to chemical bond breaking via dissociative excited states ($S_1$ in Figure 1b).[85–87]

NRR involves the complex nonradiative electronic population decay that can be observed between states of the same spin multiplicity (internal conversions) or between states with different spin multiplicity (intersystem crossings), see Figure 1c. Beyond common single crossings between excited states within the singlet manifold,[88] typical examples of complex internal conversions that do not involve large amplitude nuclear motions are decays induced by repeated crossings of regions of strong nonadiabaticity[89] or reflections[53], three-state conical intersections[90] or extended degeneracies between electronic states.[91] Intersystem crossings are driven by spin-orbit coupling, and are, thus, most often associated with transition metal complexes.[92] Nonetheless, they are also common in organic molecules[93–97], particularly in carbonyl compounds, when sulfur- or selenium-substituted,[98–101] and in nitroaromatic compounds.[102, 103] It should be noted that since spin-orbit coupling is generally relatively weak, the timescale necessary for observing significant intersystem crossing can range from few hundreds of femtoseconds to hundreds of nanoseconds.[104]In addition, the performance of simulations of intersystem crossing may depend on the strength of spin-orbit coupling, especially since it is also now established, experimentally and theoretically, that intersystem crossing can compete on similar time-scales as internal conversion, at least in geometrically unconstrained molecules with high density of states and overlapping spin manifolds.[105–107] Although several NAMD approaches have been formulated to describe intersystem crossing,[108] simulations over long timescales still remain challenging.

ESIPT reactions may occur in complex biological systems and are exploited for the development of sensors and sunscreens, among others.[109–112] For the purpose of our benchmark, ESIPT processes taking place on ultrafast timescales are of particular interest as they are generally simpler to simulate with most NAMD methods.[113] Molecules exhibiting ESIPT typically contain donor and acceptor units linked by an intramolecular hydrogen bond, allowing the proton to easily migrate upon photoexcitation, as in molecules like 2-(2-hydroxyphenyl)benzoxazole (HBO) and hydroxybenzo[h]quinolone (HBQ).[114, 115] The mechanism of this migration can occur via two distinct pathways, illustrated in Figure 1d. The small barrier is overcome thermally, after which the proton undergoes a “ballistic” type motion between the donor and the acceptor; alternatively, tunneling through the potential barrier is also possible. The proper description of tunneling, which is a classically forbidden process, requires the inclusion of nuclear quantum effects for the treatment of at least some nuclear degrees of freedom (e.g., involving protons). This can be done by either adopting a purely quantum description[116–118] or extending the phase space by employing coupled trajectories as, for example, in the ring polymer formalism.[119, 120] Tunneling, however, constitutes a challenge for NAMD methods relying solely on independent classical-like trajectories.

In summary, the phenomena highlighted in this section, i.e., ISO, DIS, NRR and ESIPT, are representative of a large variety of processes found in photophysical and photochemical applications, from the study of the interstellar medium to optoelectronics. Simulating the underlying photodynamics requires that NAMD methods are able to capture challenging features, as mentioned above. Therefore, determining the capability of NAMD methods to describe accurately these highlighted phenomena will provide valuable insights into their strengths. Since NAMD methods are inherently approximate and, as expected, may perform well for certain systems but less so for others, the representative phenomena discussed here are not meant to be used as criteria for direct rejection of methods that do not match the reference results. Instead, benchmarking against these phenomena is intended to help users select appropriate methods or identify areas for improvement, depending on the specific problem being addressed.

### B. Computational methods for NAMD

Many different NAMD methodologies have been developed over the years,[121] and in the interest of treating a comprehensive set of molecular benchmarks, the techniques developed and applied within the NAMD community should be broadly represented. Here, we briefly summarize some of these approaches related to molecular dynamics and (photo)chemical reactions, with a schematic overview being given in Figure 2. The NAMD approaches presented below are organized mainly based on the underlying assumptions, and for the sake of conciseness, we only provide some general information on the methods and on the key references.

Note, however, that we cannot provide details on all the effects that these assumptions have on the outcome of a simulation, as this is, indeed, the ultimate goal of benchmarking.

Early efforts in the field can perhaps be traced back to the development of real-space grid-based solvers for the time-dependent Schrödinger equation, such as the discrete variable representation (DVR) approach,[122, 123] or sparse-grid approaches.[24, 124] Going beyond the DVR picture whilst maintaining the concept of a fixed underlying basis, the multiconfigurational time-dependent Hartree (MCTDH) family of methods[125, 126] comprises a powerful range of exact numerical wavefunction solvers, including the original and multilayer (ML-MCTDH) formulations,[127, 128] which are available for fermions, bosons, and mixtures of the two.[129] There are also approaches within this framework to treat density operators ($\rho$-MCTDH).[130–133] While fundamentally different at first sight, the time-dependent density matrix renormalization group (TD-DMRG)[134, 135] and tree tensor network state (TTNS) extensions thereof actually are another way to solve the ML-MCTDH equations of motions.[136, 137]

A challenge for applying these approaches to high-dimensional systems is that the global electronic PESs and related couplings as well as local operators are required to be in a separable form, either sum-of-products (for MCTDH) or a tree, multilayer representation (for ML-MCTDH), as a prerequisite.[138, 139] Alternatively, additional time-dependent sparse-grid approximations can be used together with MCTDH methods.[140, 141] Furthermore, the surfaces and couplings should preferentially be in a diabatic representation to avoid numerical issues arising from singularities at conical intersections in the adiabatic representation.

Relaxing the constraint of having a fixed basis has led to dynamical wavepacket methods such as full multiple spawning (FMS)[142] and ab initio multiple spawning approaches (AIMS),[7, 143] as well as the recent variants of AIMS that have been developed to include external fields,[144] spin-orbit coupling,[145] and to optimize the computational efficiency.[146, 147] In this family of methods, nuclear trajectory-basis functions represented by frozen Gaussians evolve classically on adiabatic PESs. In addition, there is the closely related range of techniques stemming from multiconfigurational ansätze, such as the coupled-coherent states approach (CCS),[148] the multiconfigurational Ehrenfest method (MCE),[149] and the ab initio multiple-cloning algorithm (AIMC),[150] that in general use different PESs than the adiabatic ones to evolve the trajectory-basis functions.

Using a fully variational framework with Gaussian wavepackets leads to the variational Gaussian-based approaches, namely the variational multiconfigurational Gaussian (vMCG) formulation and Gaussian-based MCTDH (G-MCTDH).[151–153] More recently, direct-dynamics extensions of vMCG (DD-vMCG) have been developed enabling implementations with on-the-fly electronic structures,[154] forgoing the need to precompute a global PES.

While the above trajectory-guided methods have been developed to directly tackle the time-dependent molecular Schrödinger equation, quantum-classical methods simplify the coupled electron-nuclear quantum problem by decomposing it into a quantum electronic system coupled to a classical-like nuclear system. Quantum-classical approaches can capture some of the quantum aspects of nuclear dynamics by using an ensemble of trajectories to represent the nuclear density, and they are extremely appealing due to their tractable computational cost.

Surface-hopping methods, in their original formulation[40, 155, 156] and further developments [157–164] have become an important class of algorithms for simulating mixed quantum-classical dynamics. Of these, the fewest switches surface hopping (FSSH) approach of Tully[40] is perhaps the most popular choice in current practice. Alternative hopping formalisms have been introduced and gained popularity to circumvent the direct calculation of nonadiabatic or overlap couplings, such as Landau-Zener surface hopping (LZSH) or Zhu-Nakamura theory.[157, 165–167] Decoherence corrections (dFSSH), which are intended to cure the overcoherence problem with FSSH and ensure consistent numerical propagation of classical and quantum populations by enforcing population alignment during decoherence events,[168, 169] have been developed from numerous different approaches,[170–175] as well as alternative hopping algorithms.[165, 176–179] While surface-hopping methods are a popular choice in applications, there remain many open questions in terms of conceptual grounds[169] and formulation of an optimal algorithm, particularly, pertaining to how velocity rescaling and frustrated hop protocols are implemented, [180, 181] or concerning the treatment of trivial crossings.[182, 183] As such, a range of alternative surface-hopping approaches have also been developed, including approximate methods based on the exact factorization of the full molecular wavefunction,[184, 185] which can involve coupled[177, 186–188] or auxiliary[189–192] trajectories (SH-XF). Other FSSH-based variants that offer improved accuracy have also been developed[175, 179, 192–196]. For example, while the standard implementations of surface-hopping methods conserve the energy of each classical nuclear trajectory in the ensemble, it has been pointed out that energy should be conserved over the trajectory ensemble as a whole, as quantum-trajectory surface-hopping methods do,[195, 196] which eliminates the need for velocity rescaling and special treatments for forbidden hops. Surface hopping has also been generalized beyond internal conversion[160, 197, 198] and beyond the usual quantum-electron/classical-nuclei partitions.[199] Recently, using the semiclassical mapping formalism, a mapping approach to surface hopping (MASH) has also been introduced.[200–202]

Mean-field type approaches are another major category of trajectory-based dynamics methods. While Ehrenfest dynamics[203] belongs to the family of quantum-classical methods and is perhaps the most well-known method of this type, a number of notable improvements have been developed. Quantum-Ehrenfest (Qu-Eh) combines the idea of evolution on an average potential with quantum dynamics and has been related to a particular formulation of vMCG.[204] The ab initio multiple cloning (AIMC) has been proposed as an alternative to address the coherence issues inherent in Ehrenfest trajectories, naturally incorporating decoherence through *cloning* events.[150, 205]. A valuable improvement of semiclassical Ehrenfest was the inclusion of coherence and decoherence effects in the coherent switching decay of mixing (CSDM) method.[206–208] One important branch of these developments stems from the semiclassical initial value representation.[209–213] More recently, fully linearized,[214–218] and partially-linearized[219, 220] approaches based on the mapping formalism (linearized spin mapping, LSM) have been put forward,[221] which have also proven to offer improved accuracy over the Ehrenfest limit. More generally, it is worth noting that mapping Hamiltonian approaches, such as the Meyer-Miller model, provide an effective framework for investigating nonadiabatic dynamics by transforming discrete quantum states into continuous physical variables.[221–228] In this way, electronic and nuclear degrees of freedom can be treated on an equal footing in the phase space.[229–231] The exact factorization of the full molecular wavefunction can also be used as a starting point to derive coupled-trajectory mixed quantum-classical (CTMQC) dynamics,[232–235] such that deco-herence effects naturally emerge as correction terms to the mean-field Ehrenfest equations. Additional variants of mean-field Ehrenfest have been developed to include external fields and simulate transient absorption spectroscopy.[236, 237]

The mixed quantum-classical Liouville (MQCL) equation represents another class of NAMD methods, which adopts the density matrix formalism and rigorously introduces the classical limit starting from the quantum-mechanical Liouville-von Neumann equation.[238, 239] MQCL is exact in various conditions, for instance in the case of linear coupling between the quantum and the classical subsystems, and has been shown accurate for proton transfer and proton-coupled electron transfer reactions.[240, 241] It is important to mention here that by introducing approximations in MQCL, as shown in Ref.[162], the surface-hopping scheme can be derived, hence, putting such a phenomenological approach on a firmer ground. The MQCL has also been used to derive mean-field[242] and improved mean-field algorithms using full[231, 243] and partial linearization techniques[244, 245], introducing the groups of linearized semiclassical methods (LSC) and partially linearized methods (PLDM). A closely related range of techniques have been developed using a path integral formulation[246], which also permits full[247] and partial linearization[248] approximations.

Still in the density matrix representation, a further range of numerically exact approaches have been developed to solve the electron-nuclear problem. Real-time path integral methods such as the quasi-adiabatic path integral (QuAPI) methods,[249, 250] and recent extensions such as the quantum-classical path integral approach[251] and the small matrix path integral approach,[252] use the Feynman path integral method to propagate quantum degrees of freedom. Nonequilibrium Green's functions (NEGF),[253–255] often applied to mesoscopic systems and transport problems, can also be of interest to study proton tunneling reactions.[256] Ring polymer molecular dynamics (RPMD) methods, that extend the imaginary-time path-integral formalism to the real-time domain, have been originally developed for ground-state calculations to account for nuclear quantum effects,[257] but have been also generalized in various flavours to treat nonadiabatic problems.[258–262]

Other related quantum dynamics methods, which employ either a density-matrix or a wavefunction formalism, are also worth mentioning here, such as nuclear-electronic orbital (NEO) methods,[118, 263, 264] the hierarchical equations of motion (HEOM) method,[265, 266],

TTNS approximations[267, 268] and tensor-network-based time-evolving block-decimation techniques (TEBD), among others.[269, 270]

A careful comparison between quantum dynamics and trajectory-based NAMD should address the potential for zero-point energy (ZPE) leakage in the latter, especially in long-time-scale processes such as intersystem crossings. Recently developed Hessian-free ZPE correction methods provide a promising solution to improve the consistency of simulations across these approaches.[271, 272]

Recent, provably efficient quantum-computer (QC) algorithms for NAMD simulations present additional opportunities for benchmarking. QC NAMD algorithms use the exponentially large Hilbert space of the QC to represent the Hilbert space of the nuclei and electrons of molecules. This representation allows them to prepare states, simulate dynamics, and measure observables using time and memory polynomial in system size. QC algorithms have been developed for both analog quantum simulators and fully programmable digital quantum computers. Most of the analog approaches are simulations of vibronic-coupling models,[273–276] and have been implemented on quantum hardware to simulate dynamics around conical intersections,[274] charge transfer,[275] and photoinduced dynamics in molecules such as pyrazine.[276] Digital algorithms require large-scale, fault-tolerant quantum computers, but they could simulate the time-dependent Schrödinger equation of all nuclei and electrons on a grid exactly, up to a known and controllable error.[277–281] Other digital algorithms for near-term QCs use variational principles, but lack provable error bounds.[282, 283] QC approaches both require new benchmarks to allow fair comparisons with heuristic classical methods and deliver new tools for error-analysis which will allow developers of classical-computer algorithms to more accurately bound the errors of their simulations.

Figure 2 provides an overview of the classes of methods presented in this section. However, NAMD is a rapidly evolving field, with a wide variety of methods being constantly developed and improved, thus, it is challenging to provide an exhaustive list of all the methodologies and variations that have been introduced. Our goal is to offer an overview of the most common and widely used approaches. We also recognize that some methods on the list have primarily been used with low-dimensional model systems so far, but in principle, they could be adapted for realistic molecular systems. Figure 2 also gives an additional categorization

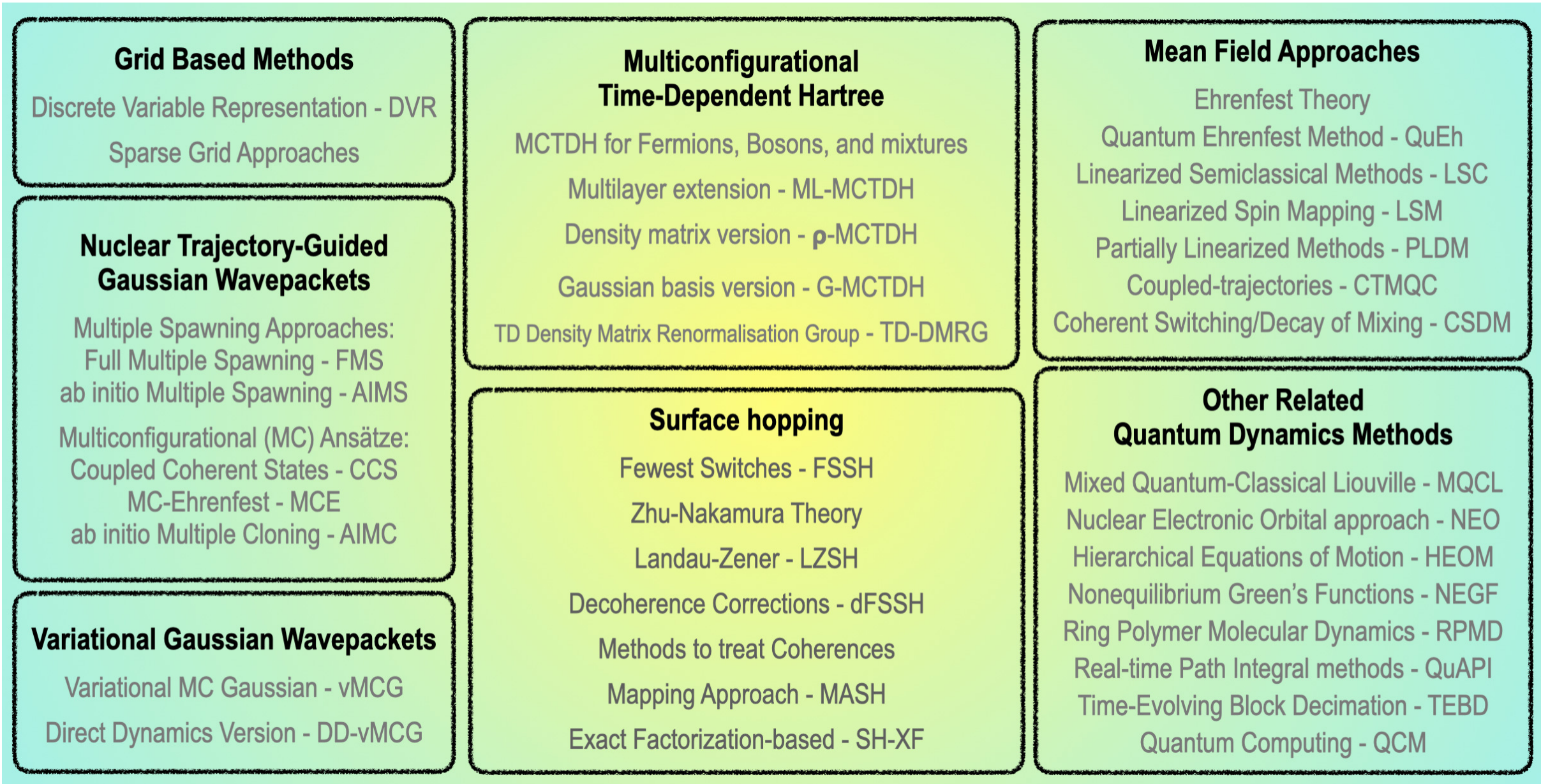

FIG. 2: Selected techniques from the range of NAMD methods that are relevant to establishing molecular benchmarks.

of the methods by organizing them in "categories" highlighted by the boxes.

### C. Electronic structure and representation of potential energy surfaces

The ingredient of an NAMD simulation that arguably plays a critical role on its outcome is the underlying electronic structure method, i.e., the level of theory at which the electronic energies, gradients, and couplings between electronic states are calculated – as highlighted in numerous studies.[106, 284–288] The impact of the electronic structure on the result of NAMD remains, however, challenging to predict. While very different PESs calculated from two different electronic-structure methods often lead to different excited-state dynamics,[289] examples in the literature show that this correlation does not always hold: vastly different PESs can lead to similar dynamics, and similar PESs can lead to different results in NAMD.[287, 290–292] In any case, it is critical to ensure that the electronic-structure quantities for any benchmark system are obtained consistently to fairly compare the outcome of the NAMD simulations.

For the current standard practice of benchmarking on low-dimensional analytical models,

ensuring consistency in electronic structure between different NAMD methods is a minor issue. A model Hamiltonian usually provides analytical expressions for energies and (diabatic) couplings, and perhaps even for gradients and for nonadiabatic couplings, making it simpler to ensure that different NAMD simulations are performed using the same electronic information.[293]

In more realistic scenarios, NAMD is often carried out with electronic structure calculated on the fly (also called direct-dynamics) using trajectory-basis functions or quantum-classical techniques. This terminology means that any electronic-structure quantity for the dynamics is calculated locally, i.e., at the current nuclear configuration at that time step, rather than being precomputed or predefined over the full configuration space. In these cases, resolving the electronic structure problem becomes a critical step before establishing benchmark systems. To meaningfully compare different NAMD methodologies and softwares, it is essential to define the level of electronic-structure theory, ensuring that the underlying electronic-structure quantities remain consistent for all NAMD methods being compared.

An ideal electronic structure method should fulfill several criteria: 1) provide electronic energies, nuclear gradients, and any required couplings (e.g., nonadiabatic couplings, spin-orbit couplings, transition dipole moments), 2) describe all electronic states involved in the dynamics with equal accuracy across the entire configuration space encountered during the dynamics, 3) be numerically robust, 4) capture the potential multiconfigurational character of electronic wavefunctions, and finally, 5) be computationally affordable.[7, 294]

Multiconfigurational methods[295] like multiconfigurational self-consistent field (MC-SCF), (state-averaged) complete active space self-consistent field (CASSCF), or complete active space configuration interaction (CASCI) are computationally expensive but include static correlation, often providing a qualitative correct picture of the PES. These methods allow the user to select the active space orbitals (occupied and unoccupied) that dominate the excited-state characters of the molecule of interest.[296] Making an informed choice of active space, that is, making it as compact as possible while still remaining sufficiently stable throughout the dynamics, can offer a good compromise between cost and accuracy. However, in many cases a CASSCF or CASCI approach may not be accurate enough, due to the lack of dynamical correlation. This can be incorporated through the application of perturbation theory (e.g. via multi-state or single-state CASPT2)[297] or with multireference methods (e.g. MRCIS or MRCISD).[295] Such methods bring an improved description of the PESs, in particular when excited electronic states of different characters interact, but also increase substantially the cost of the calculation.[297] Alternatively, scaled CASSCF methods (e.g. $\alpha$-CASSCF,[298, 299]) introduce empirical corrections to state-averaged CASSCF, improving the description of PESs while maintaining computational efficiency, and have been successfully applied to study photochemical ring-opening and isomerization reactions.[299–301] If structural rearrangements during the dynamics drive the molecule to regions of the PESs far from the Franck-Condon region, a single computationally-affordable active space might not provide enough flexibility to describe the photoproducts with the same accuracy as the initial molecule often leading to instabilities in the electronic structure. As a computationally efficient alternative, floating occupation molecular orbital complete active space configuration interaction (FOMO-CASCI)[302] was also employed in combination with NAMD.[303]

For large molecular systems, linear-response (LR) time-dependent (TD) density functional theory (DFT) is a practical alternative due to its excellent balance between cost and accuracy. However, LR-TDDFT often suffers from limitations due to its approximations necessary for practical applications.[304] One of them is its reliance on the *adiabatic approximation*, which hinders describing conical intersections with the electronic ground state, electronic states with double-excitation character, and charge-transfer transitions and Rydberg transitions; range-separated hybrids may help with these last two problems. [305–315] These shortcomings may hamper the applicability of LR-TDDFT in NAMD simulations for systems exhibiting such features. In general, for any application to a molecular system, the choice of an adequate density functional may be challenging and requires careful benchmarking.[316–318] Spin-flip variants of these methods exist,[319, 320] which can address some of these issues but often introduce spin contamination, except for spin-adapted spin-flip methods.[321–324] Some of the aforementioned limitations of LR-TDDFT can be overcome using the ensemble-DFT-based approach which combines multireference methods within a density functional theory framework.[325]. In a related approach, the mixed-reference spin-flip TDDFT (MRSF-TDDFT) technique[326] has been proposed recently. At variance with LR-TDDFT, MRSF-TDDFT was shown to predict the correct topology of conical intersections with the ground state and to describe excited states with significant double excitation character[327, 328]. Hole–hole Tamm–Dancoff approximated (hh-TDA) density functional theory[329] constitutes another variant of LR-TDDFT adequately describing conical intersections and combined with NAMD.[330] Relatedly, particle-particle RPA can describe double excitations well,[331] and, along with their oscillator strengths, related to their couplings, so can dressed frequency-dependent TDDFT.[332, 333]

We note that, in addition to conventional LR-TDDFT, real-time TDDFT (RT-TDDFT)[304] has also been used. In RT-TDDFT the electron density is propagated by integrating the time-dependent Kohn–Sham equations. RT-TDDFT (or more broadly, real-time electronic structure methods)[334] can be naturally coupled with Ehrenfest dynamics[335] to propagate nuclei classically with forces derived from a weighted average of all electronic states. However, in this approach there is no need for the explicit determination of individual electronic states and their couplings.

The algebraic diagrammatic correction to second order, ADC(2), is a wavefunction-based single-reference method that has been exploited for NAMD.[336, 337] This method, in its original implementation, possesses some limitations – it cannot describe conical intersections with the ground electronic state[284] and suffers from a systematic flaw for carbonyl-containing molecules.[338] However, its overall accuracy and efficiency in describing excited PESs and their coupling regions,[313] as well as its reliability, makes it a key contender for the NAMD of medium-sized molecular systems. ADC methods are closely related to coupled cluster (CC) methods,[339] which were historically not a popular choice for NAMD due to their intrinsic instabilities.[337, 340] However, recent CC implementations managed to resolve some of these problems, and have opened the door for CC-based NAMD simulations.[341, 342]

Semi-empirical multireference methods based on multiconfigurational configuration interaction wavefunctions built from FOMO-CI,[22, 178, 343, 344] particularly those reparameterized based on high level calculations, or the multi-reference configuration interaction based on the orthogonalization-corrected model Hamiltonian (MRCI/OMx),[22, 345] may offer an affordable alternative for describing conical intersections and complex electronic densities. If reparameterization has already been performed for the molecule of interest, these methods can be a suitable choice for benchmarking. They offer electronic structure quantities at a low cost, enabling long propagation times and large numbers of trajectories to be evolved.[186, 287, 346–349]

The use of a given electronic-structure method to benchmark on-the-fly NAMD techniques is challenging, even if all the input parameters (and initial orbitals) are provided. Ideally, the same quantum-chemical code should be used to ensure a one-to-one comparison, as minor implementation details, such as convergence criteria or algorithmic differences, can impact the final results. To promote accessibility and broader participation of community members, benchmarks should preferably employ freely available or open-source quantum-chemical codes that are widely used within the NAMD community (e.g., OpenMolcas,[350] Bagel,[351] Orca,[352] NWChem[353], GAMESS[354], MNDO,[355] MOPAC-PI[21] or PySCF[356]). For many quantum-chemical methods, ensuring consistency between two calculations is relatively straightforward if one uses the same version of a given quantum-chemical code and the same input parameters. The case is harder for multireference and multiconfigurational methods, for which it is crucial to ensure that the very same initial molecular orbitals are included in the active space. This can be achieved by making sure that the starting orbitals are provided as a wavefunction file.

Grid-based methods for quantum dynamics require integrals to be performed over the entire nuclear configuration space. On-the-fly dynamics is hard to perform for such methods (even though recent forays in this direction have been made[357, 358]), therefore they often rely on precomputed electronic structure quantities to fit or build analytical models. A very common approach for obtaining high-dimensional model potentials is to parameterize the PESs with vibronic coupling (VC) models, where the simplest form is the linear VC (LVC). [359] An LVC model proposes to build a harmonic expansion of the diabatic states around the Franck-Condon region, using information from electronic-structure calculations, along with the linear coupling among these diabatic states. While VC models can accurately capture the ultrafast decay in NAMD, in their simplest LVC form, they are limited by their underlying harmonic approximation for describing the PESs and can only be applied to relatively rigid systems. Despite these shortcomings and when used on suitable systems, the LVC approach has recently gained popularity as a cost-efficient mean for comparing differ-ent trajectory-based approaches with accurate quantum dynamics results in high-dimensional systems.[51, 54, 55, 92, 181, 360, 361] Here it is worth noting that recent advances in artificial intelligence and machine learning have significantly enhanced the accessibility of high-dimensional and anharmonic analytical potentials, reducing the computational cost of electronic structure and improving the fitting procedures,[362] thus pushing NAMD simulations to longer time scales.[363–365] In this context, benchmarking efforts will become even more critical in the future, particularly as machine learning based interatomic potentials (MLPs) evolve into widely adopted tools for NAMD. Well-defined benchmarks will be crucial not only for testing traditional electronic structure methods but also for providing a structured framework to assess how well MLPs reproduce reference electronic structure results within the same NAMD framework. Moreover, stable and reliable MLPs have the potential to revolutionize how NAMD methods are evaluated by enabling rapid and extensive testing, and facilitating the efficient exploration of the parameter space in existing NAMD techniques.

Finally, the use of fitted, analytical potentials versus on-the-fly dynamics for benchmarking needs to be addressed further. There is a clear and obvious advantage in developing models based on analytical potentials, as they directly allow quantum dynamics simulations to be performed and (near) numerically-exact solutions to be used as a reference. However, most common applications of NAMD focus on molecules for which a parametrization in full dimensionality is often inaccessible and that are therefore more easily described by on-the-fly simulations. Hence, NAMD benchmarks should be best conducted based on both approaches: fitted/analytical potentials and on-the-fly dynamics. One should stress that in general, a NAMD simulation carried out on precomputed, fitted potentials in reduced dimensionality cannot be compared with NAMD conducted with direct dynamics in full dimensionality (see for instance a quantum dynamics study on a 2D model of retinal,[42] followed by fully-dimensional direct dynamics[366] and experimental evidence[367] demonstrating the necessity of additional degrees of freedom). This is because the configuration space that can be explored is predefined in a precomputed model, constraining the dynamics to a certain region of the nuclear configuration space. It might be useful, however, to conduct simulations on model potentials using on-the-fly NAMD simulation methodologies to provide a strict comparison of their performance against accurate grid-based methods.

A plausible solution to such an issue is to fit directly a global PES into analytical separable (sum of products or tree) form. In this way, it could be used by the whole range of NAMD techniques. From a purely algorithmic perspective, one can distinguish two classes of such fitting procedures: (i) those based on machine learning (ML) or neural networks (NN) and (ii) those relying on a functional ansatz related to tensor decomposition algorithms.

The first category includes methods based on machine learning or neural networks, such as single-layered Neural Networks with specific activation functions[368] and Gaussian Process Regression with separable multidimensional kernels.[369] The second category involves methods exploiting PES smoothness under separable form constraints, including Smolyak interpolation scheme with non-direct product basis[370] and the Finite Basis Representation (FBR) family of PES representations.[371–373] FBR models can be optimized from scattered reference data and have been applied to various physico/chemical processes, including vibrational problems (6D/9D),[371, 372] reactive scattering processes (13D/15D/72D),[372, 374] and nonadiabatic dynamical problems (12D).

### D. Initial conditions for the dynamics

Any NAMD simulation requires a definition of the initial state of the molecule before being excited by light or before the dynamics is started. Therefore, a critical aspect to discuss is the nature of this initial molecular state for the molecule of interest, be that the ground state of the molecule or the state directly generated by the photoexcitation process.[375]

Following the time-dependent perturbation theory to first order for a system with two electronic states,[4] one can show that the first-order contribution to the molecular state immediately after excitation by an infinitely short pulse (a $\delta$-pulse) is simply the initial ground-state nuclear wavefunction (multiplied by the transition dipole moment between the ground and the excited electronic state). In other words, if the molecule is excited by a very short pulse, a commonly accepted approximation is to simply project the ground-state nuclear wavefunction onto the desired excited electronic state.[376] This approximation, often referred to as the sudden, or vertical, excitation, dramatically simplifies the preparation of initial conditions for NAMD, as it neglects the time duration of the excitation process (e.g. an experimental laser pulse) and the precise nature of the molecular state formed upon photoexcitation.

Within this sudden excitation, the initialization of a quantum dynamics simulation only requires the nuclear wavefunction associated with the ground electronic state for the system of interest, often taken as the ground vibrational state for all modes considered. This nuclear wavefunction can be obtained by imaginary-time propagation or, for potential energy surfaces invoking a harmonic approximation, simply from a Gaussian nuclear wavefunction. Similarly, the most commonly employed strategy for trajectory-based methods consists first of sampling an approximate ground-state distribution. The harmonic Wigner distribution, constructed from the molecular equilibrium geometry and its harmonic normal modes, is often used to sample representative initial conditions (nuclear momenta and positions).[6, 375] Once the initial ground-state nuclear wavefunction (quantum dynamics methods) or ground-state nuclear momenta + positions (trajectory-based methods) are acquired, they can be projected onto the desired excited electronic state to begin the NAMD.

While the protocol described above is the most commonly employed strategy to initialize a NAMD simulation, it relies on a series of approximations,[377] namely that (i) the molecule is in its electronic and vibrational ground state before photoexcitation and (ii) that the laser pulse employed is infinitely short (or at least short enough for its bandwidth to overlap with all necessary vibrational states in the excited electronic state for a projection of the ground-state nuclear wavefunction), meaning that a perfect nuclear wavepacket is generated in the excited electronic state(s) of interest. We note that the initialization of quantum dynamics simulations also often relies on the Condon approximation, that is, the transition dipole moment does not depend significantly on the nuclear geometries under the initial wavepacket. Nevertheless, care must be taken since cases of strong violation of Condon approximation have been reported.[378]

For floppy molecules with multiple dihedral angles and low rotational barriers, harmonic Wigner distributions are not suitable for sampling initial conditions. Improvements in the generation of the ground-state probability density can be obtained for trajectory-based methods by using *ab initio* molecular dynamics (AIMD). Initial positions and momenta can be obtained from long, equilibrated ground-state molecular dynamics simulations, providing a more accurate representation of the nuclear phase space. Incorporating zero-point energy (ZPE) in the dynamics requires the use of a quantum thermostat (QT),[379–381] as a regular 300K AIMD would lead to molecular distributions that are too narrow in comparison to their ZPE equivalent.[382] QT-AIMD can overcome some limitations of the harmonic Wigner sampling, in particular for flexible molecules with low-frequency (anharmonic) vibrational modes. Using the harmonic Wigner sampling for molecules with photoactive low-energy modes can lead to severe artifacts in the ensuing excited-state dynamics – an issue fixed with the QT-AIMD.[383]

Moving beyond the sudden excitation requires a more careful inclusion of the external electric field in the simulation, aiming for more robust comparisons with experiments. Most methods for NAMD have been extended to incorporate photoexcitation triggered by an explicit laser pulse (e.g., Refs. 31, 160, 236, 237, 384–386). This strategy, though, does appear to stretch the approximations of methods like surface hopping for longer laser pulses,[144, 387, 388] and modifications of surface hopping based on Floquet theory were presented in the literature.[389, 390] Different works have discussed photoexcitation beyond laser pulses, including incoherent sunlight[376, 391, 392] or a periodic drive,[393] in NAMD. Building the effect of a laser pulse within the initial conditions was also suggested.[84, 394, 395] Furthermore, upon initial photoexcitation by a laser pulse, a group of electronic states may be excited and the subsequent dynamics can differ depending on whether the system evolves from a superposition of states (pure) or a mixed ensemble.[396]

Another issue that needs to be addressed in the context of benchmarking different families of methods for NAMD is the representation of the electronic states. In conventional trajectory-based methods, the nuclear dynamics is usually performed by invoking electronic properties in the adiabatic representation. On the other hand, quantum dynamics methods often rely on the – more convenient – diabatic representation to avoid encountering singularities of the nonadiabatic couplings at conical intersections. Therefore, in the initialization of a quantum dynamics simulation, the ground-state nuclear wavefunction needs to be projected onto a given diabatic electronic state. For a proper assessment of trajectory-based methods against quantum dynamics results, the initial electronic diabatic state needs to be translated appropriately into an adiabatic state or a linear combination of adiabatic states when the trajectory-based simulation is performed with methods employing the adiabatic representation.

### E. Observables and properties

Benchmarking NAMD methods faces the challenge of identifying a unique, clearly defined, and quantifiable “result”. In contrast, electronic structure benchmarks are based on well-defined numerical values such as electronic energies, or optimized geometries. The outcome of a NAMD simulation involves a time-dependent molecular wavefunction, with the desired results and properties depending on the specific system and phenomenon under study. Nevertheless, we aim to identify key properties and observables that can facilitate both qualitative and quantitative comparisons across different methods for NAMD.

In this section, we use the term *observable* in the physical sense, namely a quantity that is directly determinable from an experiment and is, therefore, independent of the theoretical representation used in the calculations as it corresponds to the expectation value of an operator. It is worth mentioning here that, as it will become clear from the discussions of the next sections, in this preliminary stage of our work on NAMD benchmarks, we are not considering using experimental signals as a reference for evaluating the performance of the simulation methods. However, it is indeed likely that we will calculate experimental observables, and, in order to provide fair and unbiased grounds for comparisons, we will adopt – as far as it is possible, given the different nature of various NAMD techniques – the same methodologies to extract the experimental observables from the available simulated data. By contrast, we refer to a *property* as a quantity that may be used to interpret the simulated dynamics or an experimental measurement, but that cannot be directly measured. Below, we discuss how observables and properties can be selected for benchmarks.

For the purpose of benchmarking, observables and properties need to be selected such that different methods can be compared fairly, following several criteria. First, it is desirable that the considered observables or properties can be computed by every NAMD method under investigation. For instance, the operators involved in the calculation of expectation values should ideally have a relatively simple form to allow for the computation of high-dimensional integrals required for the quantum dynamics approaches. Similarly, it should be possible to calculate the observables directly as expectation values by reconstructing the nuclear wavefunctions. For trajectory-based methods, the observables and the properties are often calculated as trajectory averages.

Second, it is also necessary to choose a set of observables and properties that describe all aspects of the dynamical processes at play. For NAMD, this normally requires the consideration of both electronic and nuclear degrees of freedom. Different methods are unlikely to reproduce each type of observable or property equally well, so having the most diverse observable set is important for a comprehensive comparison between methods.

Finally, NAMD methods employ different electronic representations. Some electronic properties, such as electronic populations may only be accessible in or dependent on a given representation and shall be used "with care" for comparisons of methods. One way to avoid this issue is to consider representation-independent electronic properties and observables, such as optical spectra.

To make a tangible example of the observables and properties that can be of interest for understanding the processes taking place during a photochemical reaction, let us discuss Figure 3. The *absorption* of light by an organic molecule in its vibrational and electronic ground state, $S_0$ (violet Gaussian on the left) produces a photoexcited nuclear wavepacket in a singlet excited electronic state, here $S_1$ (green Gaussian on the left). The photoexcited molecule relaxes non-radiatively via *internal conversion* from $S_1$ to $S_0$, transferring the population to the electronic ground state and accessing the configurations of the various *photoproducts* (violet Gaussians in the center). The remaining contribution to the nuclear wavepacket in $S_1$ can be transferred non-radiatively to a triplet state (blue Gaussians in the center), here $T_n$, by *intersystem crossing*. Finally, the $S_1$ and the $T_n$ wavepackets can ultimately relax radiatively to the ground state (violet Gaussians on the right) via *fluorescence* and *phosphorescence*.

Based on this schematic representation of a photochemical reaction, the following properties could provide a useful way of tracking the important aspects of the dynamics. The relaxation of the photoexcited system to a lower energy electronic state can be followed via the electronic (adiabatic and diabatic) populations. While easily accessible, electronic populations are a representation-dependent quantity and therefore, excited state lifetimes in different spin multiplicities should additionally be linked to an observable that is sensitive to them. Calculating the time-dependence of the energy gap distribution between two appropriate electronic PESs would additionally quantify the motion of the nuclear wavepacket away from the conical intersection seams towards a stable minimum energy geometry on the lower-energy surface and give a quantitative picture of the electronic relaxation, in particular for NRR. For molecular systems consisting of donor and acceptor moieties, transient exciton localization can be tracked, enabling the study of intramolecular energy transfer between different chromophoric units.[397, 398] Finally, an important aspect of photochemical processes is the formation of photoproducts. This is particularly important for characterizing ISO, DIS, and ESIPT phenomena, and can be probed via the associated quantum yields, preferably computed for different excitation wavelengths. For ESIPT-related processes, it would be particularly interesting to study the mechanistic details of the proton transfer, i.e., whether it is stepwise or concerted. This can be deduced from the time-evolved nuclear probability distribution of the transferring proton or the kinetic isotope effect,[399] for example.

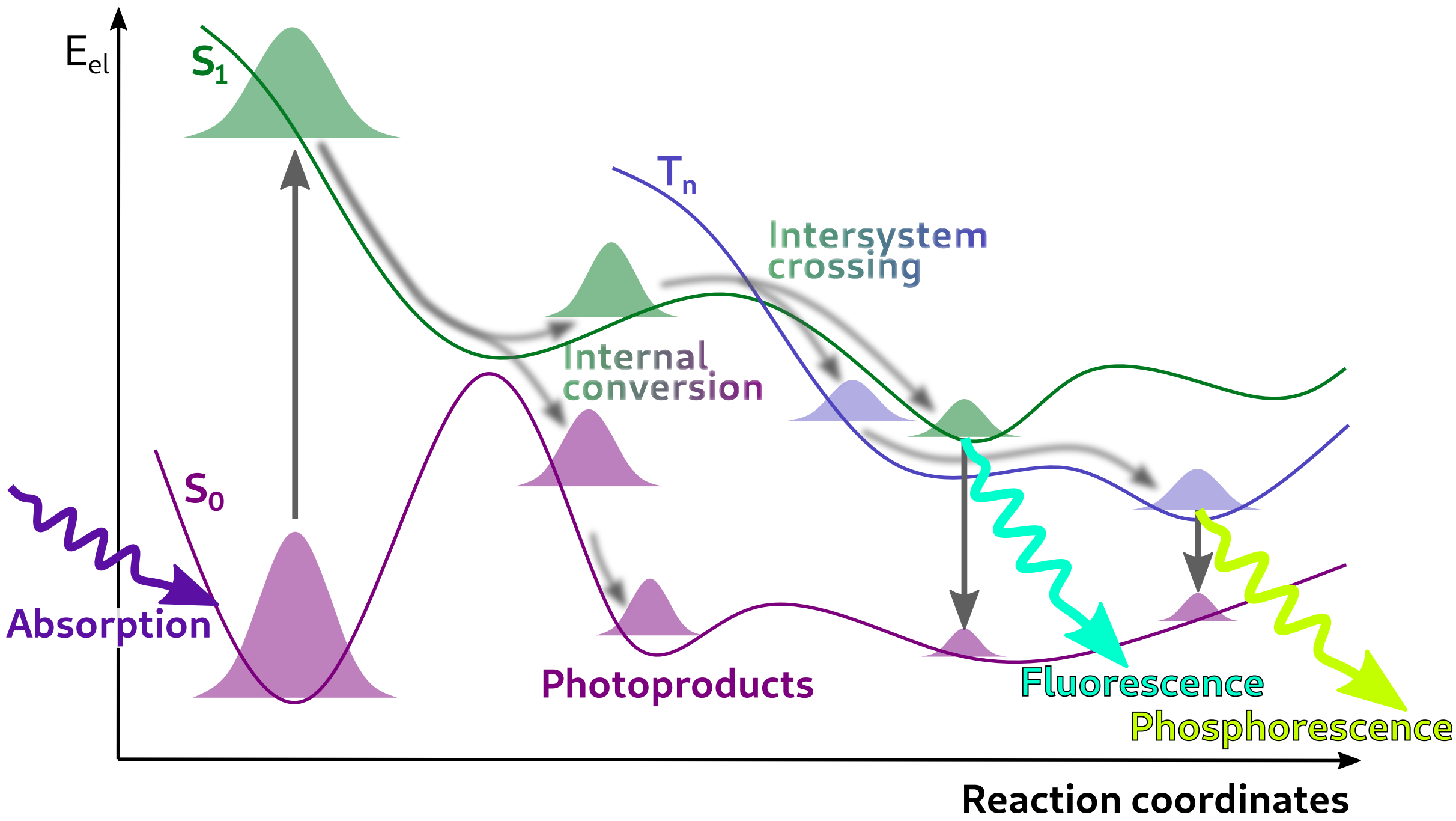

FIG. 3: Processes that can occur upon photoexcitation of a molecular system.

To properly compare with experiments, it is essential to consider various spectroscopic observables that can be used to interrogate photochemical and photophysical phenomena.[395, 400–405] Optical transient absorption,[406–409] time-resolved X-ray absorption,[410] 2D electronic spectroscopy[405, 411] and time-resolved photoelectron spectroscopy[412, 413] are several complementary techniques that directly probe the dynamical changes in electronic structure associated with a particular process. All of these techniques can, in principle, distinguish between states of different spin multiplicity,[414–416] and the method of choice may depend on whether the associated valence or core excitation spectra or the photoionization cross-sections provide the greatest contrast between the molecular species and states of interest. In particular, 2D electronic spectroscopy is an ultrafast optical technique capable of providing critical insights into coherence, which signifies the simultaneous evolution of electronic and vibrational dynamics in complex natural and synthetic systems[417] Coherence refers to the in-phase evolution of specific degrees of freedom and, in quantum mechanics, is formally described by the off-diagonal elements of the density matrix, encompassing both electronic and vibrational components.[418] Capturing such coherent phenomena in NAMD simulations[418, 419] remains a significant challenge due to the need for consistent and accurate propagation of both electronic and nuclear degrees of freedom at the surface crossings.

One advantage offered by time-resolved photoelectron spectroscopy is that processes like electronic population transfer involving an optically-dark electronic state can be directly observed.[412] Recently, attosecond transient absorption spectroscopy[420] and multiphoton ionization[421] have been used to measure and distinguish adiabatic and non-adiabatic effects in the evolution of electronic coherences. Finally, time-resolved X-ray diffraction[422–424] and ultrafast electron diffraction[425, 426] spectroscopies provide a useful way of directly probing the nuclear rearrangements of molecules in real-time and can be highlighted as effective experimental tools for investigating ISO, DIS and ESIPT phenomena. In particular, ultrafast electron diffraction has shown sufficient sensitivity to monitor the motion of light atoms like hydrogen in the context of photodissociation.[427] In general, scattering experiments are beginning to be employed to detect information beyond structural dynamics, such as electronic populations[428] or indeed the rearrangement of electrons during a reaction.[429, 430] This indicates that such experiments stand to provide comprehensive and complete information about the evolution of the molecular wavepacket.

### F. The role of experiments in benchmarking nonadiabatic molecular dynamics

Experimental observables are often regarded as the "ultimate" data for providing a reference for results obtained from quantum chemical methods. Spectroscopic techniques seem especially well-suited for providing this due to their ability to reveal quantum state information on the target system. However, there are a number of challenges when drawing comparisons between experiment and theory that need to be considered, particularly when benchmarking NAMD simulations.

Ultrafast spectroscopies are reasonably young in comparison to their static counterparts, with the earliest time-resolved optical absorption measurements being performed in the 1970s.[431] Many ultrafast techniques, such as time-resolved X-ray absorption, are even newer,[432–434] which poses a number of further challenges for using this experimental data as a reference in the benchmark of NAMD. The first of these is simply the quantity of experimental data available. While ultrafast optical techniques have seen significant hardware developments,[435] such that "all-in-one" laser and spectrometer systems are now commercially available, ultrafast optical techniques are nevertheless nowhere near as ubiquitous as standard UV-vis absorption, which is now even being performed with smartphones.[436, 437]

A more serious issue for benchmarking is the reproducibility and reliability of the experimental data, coupled with the quality of the data reporting. Although chemists and physicists are generally among the least concerned about a "reproducibility crisis" in science,[438] there have been recent reviews highlighting how, for X-ray photoelectron spectroscopy (XPS), there is a non-trivial number of papers reporting experimental data with minor errors in the collection process and a much more significant number with major issues associated with the subsequent peak fitting and data analysis procedures.[439] Growing concerns about the reproducibility problem in XPS have prompted journals[440] and the community[441, 442] to produce documents on best practices for data collection, reporting, and analysis in order to try and ensure consistent standards are maintained.

No such systematic analysis exists for the ultrafast literature, and many techniques are still sufficiently novel and challenging to perform. Hence, the research focus is still far from prioritizing systematic characterization studies. Even a brief survey of the literature, though, will reveal many inconsistencies and inadequacies in what experimental parameters are reported. For example, many papers do not report how time-zero (where the pump and probe pulses are temporally overlapping) is established or whether any wavelength calibrations for detectors have been performed and how. Often, only representative pulse parameters for the pump and probe pulses, such as pulse energies, temporal duration and central wavelengths, are reported, and no spectral information provided. While not all measurements are particularly sensitive to the excitation conditions, without this information, it becomes difficult to simulate the exact experimental conditions in excited-state dynamics simulations, particularly when an explicit pulse is included. While there are often reasons for not reporting all of this information, it is clear that it would be highly beneficial to specifically design experiments for use in theoretical benchmarking studies, where a different approach to data collection is required than for a standard photophysical experiment.[443]

It is also very important to make a clear distinction between what is the true experimental signal of measurement and what are parameters extracted from a fitting or modeling of the experimental data. For example, ultrafast spectroscopies are used to extract "lifetime" information, but the reported lifetimes are normally extracted from some kind of kinetic model with an inherent number of assumptions,[444] although there are a few notable exceptions.[445, 446] As explored more extensively in a recent review,[402] the same change in an experimental observable can arise from different physical mechanisms, and it is important to note that most spectroscopies are not directly sensitive to the population dynamics, but rather to the population dynamics convoluted with a transition probability. Evaluating the trustworthiness of models and fits can be as challenging as assessing the quality of experimental data and a difficult task without direct engagement with experts.

## III. OUTLOOK – A ROADMAP FOR MOLECULAR BENCHMARKS IN NAMD

The multifaceted nature of NAMD has, to date, hindered systematic efforts toward designing molecular benchmarks, with only a few notable studies making headway.[49, 50, 53] The intention behind this Perspective is to narrow down the multitude of available choices of "benchmarkable phenomena" and encourage collaborative efforts within the NAMD com-munity towards these goals. This Section summarizes key considerations for developing molecular benchmarks and presents an executive outline for their implementation. The steps discussed in this Perspective form the *roadmap* towards a community-driven development of benchmarks for NAMD methods (Figure 4).

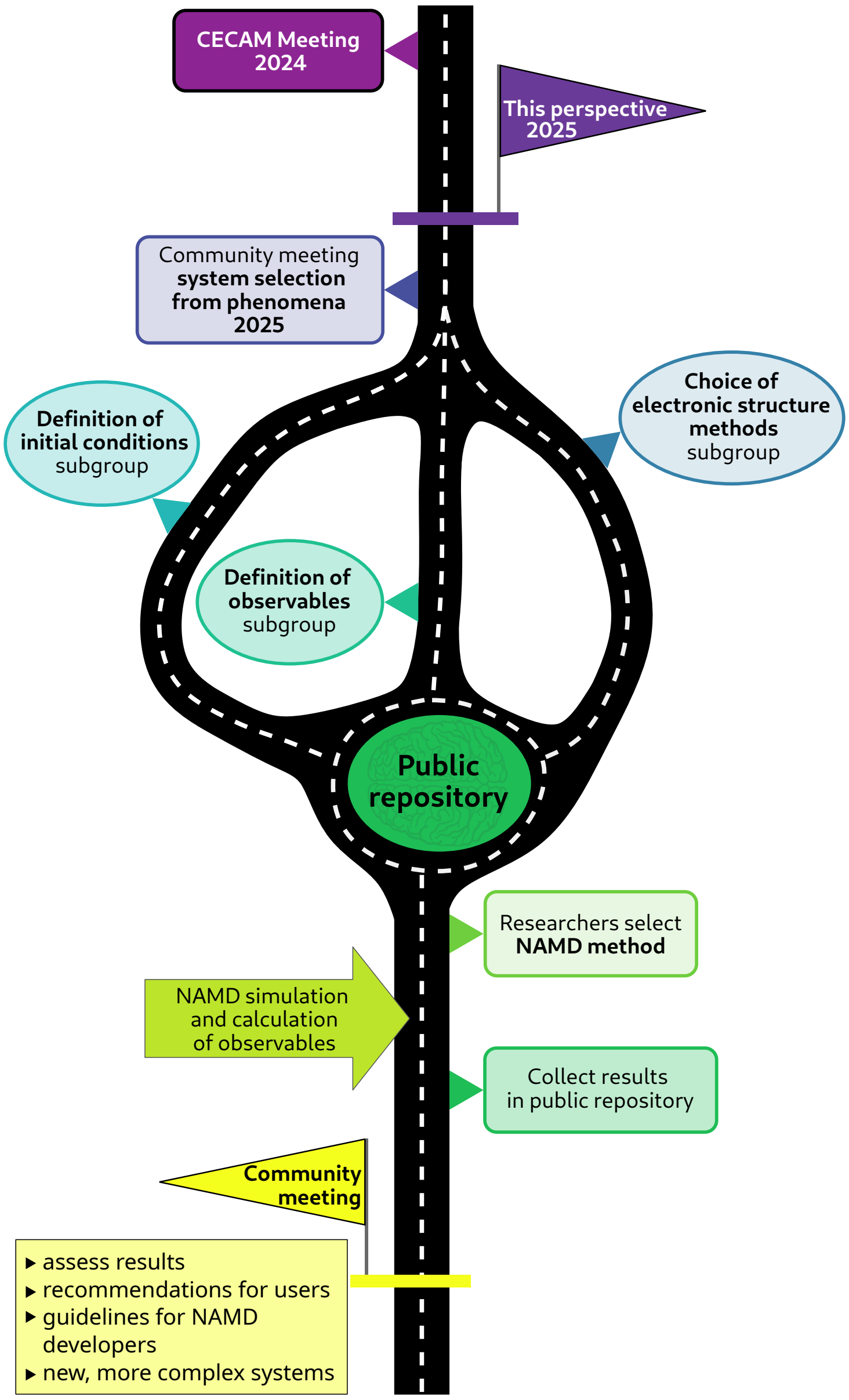

FIG. 4: Schematic representation of the steps discussed in this *roadmap* towards the creation of a benchmark set for NAMD methods and the future steps planned by the community.

First, we briefly overview key insights from the workshop in Paris,[39] which established foundational ideas for this initiative. One challenge identified by the community of NAMD users and developers is the breadth and complexity of NAMD simulations, demanding careful selection criteria to balance feasibility with scientific relevance. To ensure meaningful assessments, molecular benchmarks are meant to capture realistic nonadiabatic phenomena while minimizing reliance on model potentials with reduced dimensionality. The emphasis on realistic processes – those measurable in experiments – reflects the goal of creating benchmarks that will increase confidence in the predictive power of current theoretical developments. We have identified four relevant, although not exhaustive, groups of molecular phenomena for detailed exploration: ISO, DIS, NRR and ESIPT. They are connected to a range of observables, which will be defined early on but whose calculation represents almost the final step of the roadmap (Figure 4). Prioritizing direct calculations of molecular observables reduces the influence of different PES representations, ensuring that benchmarks maintain scientific relevance by predicting measurable quantities.

As discussed above, this relates to the question about the role of experimental data in benchmarking theoretical methods. While reproducing experimental measurements remains a key objective for the community, the direct prediction of experimental observables should currently be viewed as a goal rather than a strategy for benchmarking. Confidently bridging the gap between experiment and theory requires active collaborations from both sides in order to drive the development of all aspects and ingredients of NAMD, and as such, remains an ongoing area of research. However, this should not impede the evaluation of NAMD methods in the context of benchmarking. Therefore, we currently do not recommend the systematic assessment of theoretical results by direct comparison with experimental measurements unless the same experimental observable is calculated. Even then, caution is required. In trajectory-based NAMD methods, observables such as photoelectron spectra may be reproduced with comparatively few trajectories,[447, 448] while branching ratios require more.[106] As a consequence, it may be misleading to judge convergence and overall accuracy on the basis of a single observable. In addition, in Section II F, we focused our attention on the role of ultrafast spectroscopy in the context of benchmarking NAMD. Alternative and complementary experimental techniques, such as time-resolved mass spectrometry, can provide insights into ultrafast structural dynamics with femtosecond time resolution[449–453] (thus, having the potential to serve as a high-throughput data source for nonadiabatic simulation benchmarks), but accurate simulations of ultrafast processes in gas-phase ion (as compared to neutral) chemistry remain relatively underexplored.[454–456]

Molecular benchmarking should rather focus on comparing the theoretical approximations directly with exact or nearly exact solutions from theory, an approach that allows full control over external parameters and ensures a fair comparison. Recognizing that nearly exact solutions may not always be available or feasible to estimate, we refer to the concept of *benchmarking by comparison*. This involves comparing different methods without an absolute reference point. Even in such cases, we can establish a theoretical best estimate (TBE): a prediction from a method that provides the highest level of accuracy in treating nonadiabatic effects in that system. In this context, it is crucial that NAMD calculations are properly converged to the limit of their nuclear basis set or number of trajectories to ensure reliable comparisons. It is worth noting that TBEs have also been used in electronic structure benchmarks and have been updated progressively over time.[34, 457, 458]

Using consistent electronic structure methods and equivalent initial conditions for comparing NAMD methods is generally less contentious as an idea, but the practical implementation poses challenges. We have thoroughly discussed these complexities to address all underlying nuances. The systems of interest should avoid prohibitively expensive electronic structure, so as to allow the sufficient convergence of dynamics results, especially for the more computationally demanding approaches. Full dimensional LVC models, which require modest computational resources, may serve as a suitable testing ground for a wide range of methods, particularly in cases like internal conversion and intersystem crossing in NRR, which do not necessarily involve large-amplitude motions (as in ISO) or bond breaking (as in DIS and ESIPT). The exploration of ISO, DIS, and ESIPT phenomena requires more efforts on the accurate electronic structure evaluation. This would involve selecting electronic structure methods that are affordable, numerically stable, and widely accessible to the community through preferably software packages that are free of charge for the scientific community. Alternatively, one can take advantage of analytical potentials that have been previously obtained and used in NAMD calculations, keeping in mind that it is sometimes not straightforward to transfer a given analytical expressions into the specific form, as for instance a sum-of-products form.

In the generation of initial conditions, using simple approximations like the sudden approximation and the use of a harmonic approximation to describe the ground-state potential could mitigate the complications with initiating different NAMD methods. Nonetheless, translation between the adiabatic and diabatic representations of electronic states requires more meticulous considerations.

Research activities that are currently prioritized involve testing preliminary molecular systems and phenomena selected from a short list with strong potential as effective benchmarks. In this context, "system" refers not only to the molecule itself but also to its electronic structure representation and the possibility of calculating relevant observables. Some of the molecular systems under investigation leverage the advantages of vibronic coupling potentials. At the same time, others already have available pre-constructed PESs that would also be suitable for efficient on-the-fly simulations. Following initial scrutiny, the most promising systems will be selected for benchmarking.

Another ongoing effort involves the creation of a common online repository. Making research data openly available encourages wide participation by researchers from the field, promoting transparency, accessibility, and collaboration. An accessible online repository will allow storage of essential data such as input files, all information relevant for reproducibility and the collection of the results of the benchmarking. Here, the utilization of data science and machine learning techniques, which are rapidly advancing and increasingly permeating chemical and materials sciences, can play a critical role in supporting and maintaining data repositories. These approaches enable efficient data curation, analysis, and debugging, and are especially valuable for integrating data of varying fidelity and origin, thereby enhancing the robustness and usability of complex datasets.[459]

Once the initial set of benchmark systems is finalized and agreed upon by community members, a common set of initial conditions will be established and made available in the repository. Hereby, we recommend entrusting the preparation of initial conditions to a single dedicated research team, making sure that all types of NAMD methods are covered comprehensively. At this stage, establishing standardized input and output data formats is also anticipated to enhance the broader usability of the benchmark set. An appropriate electronic structure method, along with freely available software, will be selected for on-the-fly dynamics. All necessary quantities, input data, and (if required) initial wavefunction files will be incorporated into the repository. For methods that require analytical potentials, these will either be sourced from existing literature or parametrized and shared in the repository. Additionally, a set of relevant observables will be identified for each system, chosen to capture and represent the key aspects of nuclear and electronic dynamics. Comprehensive instructions and materials detailing the calculation of these observables will be provided to ensure consistent evaluation across all NAMD simulations. Additionally, it could be useful to standardize the tools used for post-processing and analysis of NAMD data. Selecting a dedicated Python framework[460] (or equivalent) would ensure consistency in analyzing observables, trajectory-based statistics, and error quantification, streamlining testing workflows, and promoting transparency across different benchmarking studies. The selection of initial conditions, of the electronic structure method and of the relevant observables will be performed in parallel, as indicated in the roadmap of Figure 4, ultimately converging in the creation of a repository.

Using the system information gathered in the repository, all researchers interested in participating in the benchmarking effort can test their NAMD methodologies and software on the designated test-set. This benchmarking initiative aims to engage researchers with diverse expertise, encompassing the full range of NAMD methods, from trajectory-based approaches to quantum wavepacket-propagation techniques and quantum-computing approaches. This diversity is particularly desired in the realm of *benchmarking by comparison*, as each method should ideally be leveraged to its utmost potential, using the optimal choice of parameters associated with best practices for each NAMD approach. The calculated results, along with the best practice procedures, are expected to be published in conventional research articles as well as the data shared through the repository – ensuring that benchmarks remain valuable long-term resources. The benchmarking results will be evaluated by the community members during a collective meeting. Based on the data, the goal is to assess the quality of different NAMD approaches for various molecular groups. Additionally, guidelines will be introduced for future NAMD method development, defining standardized tests and expected results to evaluate the performance of new methods.

In the long run, we foresee a continuous refinement of molecular benchmarks aligned with the ongoing advancements in the field that invariably present new challenges. More complex features will be gradually introduced, and likewise, benchmarks will be expanded to include complex systems and phenomena. This covers, for example, molecules in realistic environments (such as solvents, surfaces and materials),[461–465] an explicit treatment of light-matter interactions,[377] high density of states, and long dynamics. The utilization of machine learning approaches in these efforts has shown great promise in significantly reducing computational cost without compromising numerical accuracy, as clearly demonstrated by early studies in the field.[365, 466, 467] Finally, a synergistic approach that integrates theory and experiment (within the context of benchmarking) will inevitably emerge as a key task for the broader NAMD community.

Alongside this *roadmap*, which serves as an initial effort to disseminate our thoughts about benchmarking methods for NAMD, we aim at promoting the broader participation from community members beyond the present core group of contributors, as well as organizing regular meetings and progress reports to ensure the successful accomplishment of the plan.

## ACKNOWLEDGMENT

We are grateful to the Centre Européen de Calcul Atomique et Moléculaire (CECAM) for funding for the workshop which inspired this Perspective. We also thank the staff at the CECAM Headquarters at EPF-Lausanne, Switzerland, and at the CECAM node based in Île-de-France (CECAM-FR-MOSER), for their support in developing and advertising the program, and for their assistance in the organization of the workshop.

We acknowledge fruitful discussions and contributions from Jorge Alonso De La Fuente, Liudmil Antonov, Davide Barbiero, Lou Barreau, Julien Eng, Dario Frassi, Scott Habershon, John Herbert, Christine Isborn, Yorick Lassmann, Ben Levine, Hiroki Nakamura, Igor Schapiro, Burkhard Schmidt, Jiří Vaníček and Sergei Yurchenko.

B. F. E. C. and G. A. W. acknowledge support from the EPSRC under grants EP/V026690/1 and EP/X026973/1. B. F. E. C. also acknowledges funding from the European Research Council (ERC) under the European Union's Horizon 2020 research and innovation programme (Grant agreement No. 803718, project SINDAM) and EPSRC grant EP/Y01930X/1.

R. I. was supported by the Engineering and Physical Sciences Research Council (Grant No. EP/X031519/1).

A. K. and A. F. were supported by the Cluster of Excellence "CUI: Advanced Imaging of Matter" of the Deutsche Forschungsgemeinschaft (DFG) (EXC 2056, Project 390715994).

J. R. M. acknowledges support from the Alexander von Humboldt Foundation.

D. A. acknowledges funding from the National Science Centre (Poland) under the project No. 2022/47/P/ST4/01418, and the European Union's Horizon 2020 research and innovation programme under the Marie Skłodowska-Curie grant agreement No. 945339.

M. B. received support from the French government under the France 2030 investment plan as part of the Initiative d'Excellence d'Aix-Marseille Université (AMX-22-IN1-48) and from the European Research Council (ERC) Advanced Grant SubNano (Grant agreement 832237).

E. C. E., P. S., A. T., and I. K. acknowledge support by Google, by the Australian Research Council (FT230100653), and by the United States Office of Naval Research Global (N62909-24-1-2083).

N. D. acknowledges the support by the Croatian Science Foundation under the project numbers [HRZZ-IP-2020-02-9932 and HRZZ-IP-2022-10-4658].

J. J. thanks the Czech Science Foundation for the support via grant number 23-07066S.

A. K. acknowledges funding from the Engineering and Physical Sciences Research Council grants EP/V006819/2, EP/V049240/2, EP/X026698/1, EP/X026973/1, the U.S. Department of Energy, Office of Science, Basic Energy Sciences, under award number DE-SC0020276, and the Leverhulme Trust *via* grant RPG-2020-208.

H. R. L. was supported by the U.S. Department of Energy, Office of Science, Office of Basic Energy Sciences, CPIMS program, under Award DE-SC0024267 (ML-MCTDH and PES representations), by the U.S. National Science Foundation under grant no. CHE-2312005 (connecting ML-MCTDH to TD-DMRG), and by the American Chemical Society Petroleum Research Fund via grant no. 67511-DNI6 (electronic structure).

D. P. and Z. Q. gratefully acknowledge the project UDOPIA and the French National Research Agency (ANR), as well as Institut DATAIA Paris-Saclay, the AI institute of the Univer-sity Paris-Saclay for his Ph.D. funding (https://www.dataia.eu/). D. P. also acknowledges ANR-DFG financial support for the QD4ICEC project Grant number ANR-22-CE92-0071-02.

N. M. acknowledges the National Science Foundation (NSF) under Award No. CHE-2154829.

P. R. would like to acknowledge the School of Chemistry at the University of Nottingham, and the Hobday Fund for their start-up funding.

S. T. and S. F. A. acknowledge the Center for Integrated Nanotechnology (CINT) at Los Alamos National Laboratory (LANL), a U.S. DOE and Office of Basic Energy Sciences user facility.

F. A. and P. S. acknowledge financial support from the French Agence Nationale de la Recherche via the project STROM (Grant No. ANR-23-ERCC-0002). F. A. and L. MM acknowledges financial support from a public grant from the Laboratoire d'Excellence Physics Atoms Light Matter (LabEx PALM) overseen by the French National Research Agency (ANR) as part of the "Investissements d'Avenir" program (reference: ANR-10-LABX-0039-PALM). F. A. also acknowledges financial support from the French Agence Nationale de la Recherche via the project Q-DeLight (Grant No. ANR-20-CE29-0014).

A. P. acknowledges the support by the Croatian Science Foundation under the project number HRZZ-IP-2020-02-7262.

---

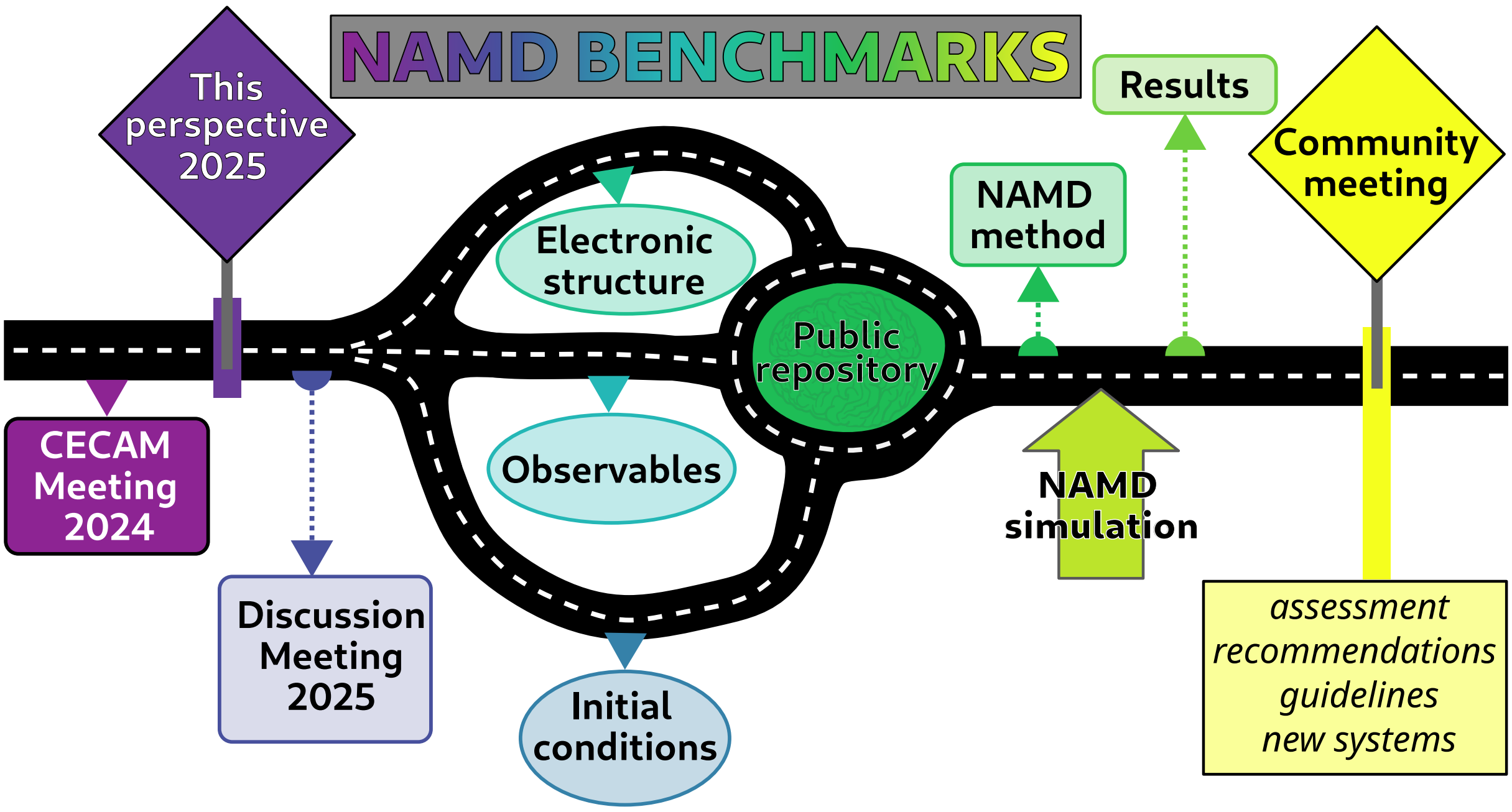

FIG. 5: TOC Graphic